\documentclass[onecolumn,prd,showpacs,preprintnumbers,amsmath,amssymb,floatfix]{revtex4}

\usepackage{graphicx}

\usepackage{bm}
\usepackage{amsfonts}
\usepackage{lineno,hyperref}
\usepackage{array}
\usepackage{microtype}
\usepackage{float}
\usepackage{epstopdf}
\usepackage{mathrsfs}
\usepackage{color}

\begin{document}
	
\title{Gravastar under the framework of Braneworld Gravity}

\author{Rikpratik Sengupta}
\email{rikpratik.sengupta@gmail.com}
\address{Department of Physics, Government College of Engineering and Ceramic Technology, Kolkata 700010, West Bengal, India}

\author{Shounak Ghosh}
\email{shnkghosh122@gmail.com}
\address{Department of Physics, Government College of Engineering and Ceramic Technology, Kolkata 700010, West Bengal, India}
	
\author{Saibal Ray}
\email{saibal@associates.iucaa.in}
\affiliation{Department of Physics, Government College of Engineering and Ceramic Technology, Kolkata 700010, West Bengal, India}

\author{B. Mishra}
\email{bivudutta@yahoo.com}
\address{Department of Mathematics, Birla Institute of Technology and Science-Pilani, Hyderabad Campus, Hyderabad 500078, India}

\author{S. K. Tripathy}
\email{tripathy\_ sunil@rediffmail.com}
\address{Department of Physics, Indira Gandhi Institute of Technology, Sarang, Dhenkanal 759146, Odisha, India}

\begin{abstract}
Gravastars have been considered as a serious alternative to black holes in the past couple of decades. Stable models of gravastar have been constructed in many of the alternate gravity models besides standard General Relativity (GR). The Randall-Sundrum (RS) braneworld model has been a popular alternative to GR, specially in the cosmological and astrophysical context. Here we consider a gravastar model in RS brane gravity. The mathematical solutions in different regions have been obtained along with calculation of matching conditions. Various important physical parameters for the shell have been calculated and plotted to see their variation with radial distance. We also calculate and plot the surface redshift to check the stability of the gravastar within the purview of RS brane gravity.	



\end{abstract}

\maketitle
	
\section{Introduction} \label{sec1}

There are lots of debates regarding the final state of a stellar collapse in astrophysics just like the initial state of the universe in cosmology. Einstein's General Relativity (GR) has been tested quite rigorously through observational results in both astrophysics and cosmology considering intermediate energy phenomena. However, in situations with extremely high energies, like initial state of the universe and end-state of stellar gravitational collapse, due to the fact that huge amount of energy is being confined in a microscopic volume, the energy density almost diverges leading GR to predict singularities where the field equations of GR break down completely~\cite{Hawking1}, leaving room for considering quantum effects to come into play to avoid the undesirable singularities. In case of stellar collapse end-states, the consideration of quantum field effects in classical GR leads to many interesting additional features about the most popular end-state solutions of black holes (BH), like emission of Hawking radiation from the event horizon~\cite{Hawking2} (which may be thought of as the boundary of a BH), but it cannot remove the singularities in this particular solution of the Einstein's field equations (EFE's)~\cite{Birrell}. The singularity occurring at the Schwarzschild radius $R=2GM~(c=1)$ is not a physical singularity as the curvature invariants remain finite here and the singularity can be removed by a co-ordinate transformation. However, the central singularity occurring at $r=0$ being a physical singularity is irremovable.

In 2001, Mazur and Mottola (MM)~\cite{Mazur1} came up with the idea of a gravitational condensate star or $gravastar$ as an alternative to a black hole, which they further developed in 2004~\cite{Mazur2}. Chapline et al.~\cite{Chapline1,Laughlin,Chapline2} by taking quantum effects into consideration have proposed that the horizon may be considered as the critical surface of a gravitational phase transition with the interior balancing the gravitational collapse of the surface by holding an Equation of state (EOS) of the form $p=-\rho$~\cite{Gliner}, where the negative pressure leading to a repulsive effect. By considering the fact that there is a phase transition at the horizon, MM extended 
this idea to the quantum fluctuations which dominate the temporal and radial components of the energy-momentum tensor at the horizon and grows large enough to lead to an EOS of the form $p=\rho$. This type of EOS is on the verge of violating causality and the last
extreme allowed and leads the interior to develop a gravitational Bose-Einstein condensate (BEC). Thus, the critical surface is replaced by a shell of stiff fluid introduced first by Zeldovich~\cite{Zeldo1,Zeldo2}. This third region is the exterior which is pressureless and has zero energy density.

The gravitational force is weaker than the other three natural forces - the strong and weak nuclear forces, and the electromagnetic force. 
This is known as the hierarchy problem in particle physics. In an attempt to solve this problem, RS proposed their first brane-world model~\cite{Randall1} (RS-1) consisting of a positive and a negative tension brane with the former brane representing our universe. The $(3+1)$-branes are embedded in a higher dimensional bulk. Only the force of gravity has accessibility to the bulk while the other three forces are confined to the brane, thus making gravity the weakest of the forces. The higher dimensional gravity is the actual gravity at it's full strength and cannot be realized in the brane. Later on, they proposed another model~\cite{Randall2} (RS-2) by sending the negative tension brane off to infinity. In this model, at low energy limit the Newtonian gravity can be recovered. 

The single brane RS-2 model has been used extensively to study cosmological as well as astrophysical problems. The modifications due to RS-2 brane gravity (BG) has been studied in the cosmological context~\cite{Binetruy,Maeda,Maartens,Langlois,Chen,Kiritsis,Campos} whereas the study on modifications due to BG in astrophysical context were initially confined mostly to study of the exterior solutions~\cite{Germani,Deruelle,Wiseman1,Visser,Creek}. However, in the interior where the gravitational collapse takes place, the brane corrections to GR should be more significant as the higher energy is involved in the collapse process~\cite{Pal,Dadhich,Bruni,Govender,Wiseman2}.

Even in the context of GR, tackling exact interior solutions for spherically symmetric matter distributions is extremely difficult~\cite{Kramer}. In the case of a braneworld, the field equations have non-locality and non-closure properties due to the presence of projected Weyl tensor term on the brane~\cite{Shiromizu} which makes it even more difficult to obtain exact interior solutions, only for the exception of uniform stellar-matter distributions and can be thought of as an idealized situation. A better understanding of bulk geometry and brane-embedding properties is required for constructing exact interior solutions with realistic non-uniform distributions which has been achieved through an elegant technique called the Minimum Geometric Deformation approach developed by Ovalle in a series of papers~\cite{Ovalle1,Ovalle2,Ovalle3}. The approach has also been applied successfully to obtain exact interior solutions for non-uniform spherically symmetric matter distributions~\cite{Ovalle4,Ovalle5,Ovalle6}.

In this paper we consider a gravastar in a RS-2 brane-world model. Such a problem has been discussed by Banerjee et al.~\cite{Banerjee1} but considering conformal motion thus freezing one of the metric potentials. However, we shall not consider conformal motion in our approach to obtain explicit solutions of the EFE's. To note that there is an effective cosmological constant on the brane but no charge has been considered in this work unlike Ghosh et al.~\cite{Ghosh2017} who have considered the problem of a charged gravastar in higher dimensions. 

The present investigation has been organized as follows: the field equations for a spherically symmetric metric on a RS-2 brane have been provided in Sec.~\ref{sec2} along with the explicit mathematical solutions to the field equations considering the EOS for the interior, shell and exterior of the gravastar whereas in Sec.~\ref{sec3} we have studied various physical parameters of the gravastars. In Sec.~\ref{sec4} the boundary conditions are computed which is followed by discussions and conclusions of the results in Sec.~\ref{sec5}.

\section{Mathematical formalism and solutions} \label{sec2}

\subsection{Field Equations on the Brane}
The EFE on the RS-2 with 3-brane has the form
\begin{equation} \label{eq1}
G_{\mu\nu}=T_{\mu\nu}+\frac{6}{\sigma} S_{\mu\nu}-E_{\mu\nu},
\end{equation}
where we have chosen $8\pi G =1$. 

The second and third terms on the R.H.S. of the above equation represent the local and non-local corrections to GR due to brane effects,
respectively. The term $S_{\mu\nu}$ is quadratic in energy-momentum arising due to high-energy effects and $E_{\mu\nu}$ represents the projected Weyl tensor on the brane which can be said to be the Kaluza-Klein (KK) corrections. These terms can be expressed as follows:
\begin{equation} \label{eq2}
S_{\mu\nu}= \frac{TT_{\mu\nu}}{12}-\frac{T_{\mu\alpha}T_{\nu}^{\alpha}}{4}+ \frac{g_{\mu\nu}}{24}(3T_{\alpha\beta}T^{\alpha\beta}-T^2),
\end{equation}

\begin{equation} \label{eq3}
E_{\mu\nu}=-\frac{6}{\sigma}[Uu_{\mu} u_{\nu}+Pr_{\mu}r_{\nu}+h_{\mu\nu}(\frac{U-P}{3})].
\end{equation}

The usual $4D$ energy-momentum tensor on the 3-brane is given as
\begin{equation} \label{eq4}
T_{\mu\nu}=\rho u_{\mu} u_{\nu}+ ph_{\mu\nu},
\end{equation}
where $h_{\mu\nu}=g_{\mu\nu}+u_{\mu} u_{\nu}$ is the projected metric on the brane, $u_{\mu}$ denotes 4-velocity, $r_{\mu}$ denotes projected radial vector, $\rho$ and $p$, respectively, denotes the energy density and pressure of the matter distribution and $U$ and $P$ denote the bulk energy density and bulk pressure respectively. The last two quantities are taken to be related by the bulk EOS $P=\omega U$~\cite{Castro} whereas the energy density of the brane and bulk are considered to be given by $U=A\rho+B$~\cite{Banerjee2}.

For a static, spherically symmetric line-element describing matter distribution on the 3-brane is given by
\begin{equation}\label{eq5}
ds^2=-e^{\nu(r)}dt^2+e^{\lambda(r)}dr^2+r^2(d\theta^2+sin^2\theta d\phi^2).
\end{equation}

The EFE on the brane, given by Eq. \ref{eq1} can be computed to be
\begin{eqnarray}
&&e^{-\lambda}\left(\frac{\lambda'}{r}-\frac{1}{r^2}\right)+\frac{1}{r^2}
 =\left[\rho(r)  \left( 1+\frac {\rho(r) }{2 \sigma} \right) +{\frac {6 U}{\sigma}}\right],\label{eq6}\\
&&e^{-\lambda}\left(\frac{\nu'}{r}+\frac{1}{r^2}\right) -\frac{1}{r^2},
 =\left[p \left( r \right) +{\frac {\rho \left( r \right)  \left( p \left( r \right) +\frac{\rho(r)}{2} \right)}
{\sigma}}+{\frac {2U}{\sigma}}+{\frac {4 P}{\sigma}}\right],\label{eq7}\\
&&e^{-\lambda}\left[\frac{\nu''}{2}-\frac{\lambda' \nu'}{4}+\frac{\nu'^2}{4}+\frac{\nu'-\lambda'}{2r}\right] = \Bigg[p(r)+{\frac{\rho(r) \left(p(r)+\frac{\rho(r)}{2}\right)}{\sigma}}+{\frac{2U}{\sigma}}-{\frac {{2 P}}{\sigma}}\Bigg].\label{eq8}
\end{eqnarray}

As evident from Eqs. (\ref{eq7}) and (\ref{eq8}), the pressure is not the same in the radial and transverse directions and hence there is a pressure anisotropy amounting to $\frac{6}{\sigma}P$, which automatically justifies the claim by Cattoen~\cite{Cattoen} that gravastars must have anisotropic pressures, without forcing the anisotropy {\it apriori} by hand. This is an essential intrinsic feature of braneworld gravastar which is absent within the framework of GR.

However, the conservation equation on the brane reads the same as in GR
\begin{equation} \label{eq9}
\frac{dp}{dr}=-\frac{1}{2}\frac{d\nu}{dr}(p+\rho).
\end{equation}

\subsection{Interior solution}
As already mentioned, the interior of the gravastar has an EOS of the form $p=-\rho$. Such an EOS is responsible for a force to be created in the interior, which is now a gravitational BEC after the phase transition occurring at the horizon (replaced by a shell for a gravastar) and is acting along the radially outward direction to oppose the collapse to continue. Plugging this EOS in the conservation equation Eq. (\ref{eq9}), we get that $p=-\rho=-\rho_c$, where $\rho_c$ is the constant interior density, thus implying constant pressure. In order to compute the metric potentials we need to replace the pressure and energy density on R.H.S of Eqs (\ref{eq6}) and (\ref{eq7}) by the pressure and energy density of the interior, respectively. This gives us the field equations in the following forms:
\begin{equation} \label{eq10}
e^{-\lambda}\bigg(\frac{\lambda^{\prime}}{r}-\frac{1}{r^2}\bigg)+\frac{1}{r^2}=
\bigg[\rho_c\bigg(1+\frac{\rho_c}{2\sigma}\bigg)+\frac{6}{\sigma}(A\rho_c+B)\bigg],
\end{equation}

\begin{equation} \label{eq11}
e^{-\lambda}\bigg(\frac{1}{r^2}+\frac{\nu^{\prime}}{r}\bigg)-\frac{1}{r^2}=
\bigg[-\rho_c\bigg(1+\frac{\rho_c}{2\sigma}\bigg)+\frac{2}{\sigma}(A\rho_c+B)+\frac{4\omega}{\sigma}(A\rho_c+B)\bigg].
\end{equation}

From Eq.~(\ref{eq10}), the first metric potential can be computed to be
\begin{equation} \label{eq12}
e^{-\lambda}=1-\bigg[\frac{\rho_c}{3}\bigg(\frac{2\sigma+\rho_c}{2\sigma}\bigg)+\frac{2(A\rho_c+B)}{\sigma}\bigg]r^2+\frac{c_1}{r}.
\end{equation}

To make the solution regular at the origin one can demand for $c_1=0$ and so we are left with
\begin{equation} \label{eq13}
e^{-\lambda}=1-\bigg[\frac{\rho_c}{3}\bigg(\frac{2\sigma+\rho_c}{2\sigma}\bigg)+\frac{2(A\rho_c+B)}{\sigma}\bigg]r^2.
\end{equation}

Now, using Eqs.~(\ref{eq13}) and~(\ref{eq6}), the second metric potential is given by
\begin{equation} \label{eq14}
e^{-\nu}=C \bigg\{\bigg(\rho_c(1+\frac{\rho_c}{2\sigma})+6\frac{A\rho_c+B}{\sigma}\bigg)r^2-3\bigg\}^{\bigg[1+\frac{\frac{6(A\rho_c+B)(1+2\omega)}
{\sigma}-3\rho_c(1+\frac{\rho_c}{2\sigma})}{\rho_c(1+\frac{\rho_c}{2\sigma})+\frac{6(A\rho_c+B)}{\sigma}}\bigg]},
\end{equation}
where $C$ is an integration constant.

From the above solutions it can be noticed that the interior solutions have no singularity and thus the problem of the central
singularity of a classical black hole can be averted.

\subsection{Active gravitational mass $M(R)$}
One can calculate the active gravitational mass for the interior of the gravastar as follows
	\begin{equation} \label{eq15}
	M(R)= 4\pi\int_{0}^{R}\rho^{eff}r^2dr=\frac{4\pi}{3}\bigg[\rho_c\bigg(1+\frac{\rho_c}{2\sigma}\bigg)+\frac{6}{\sigma}(A\rho_c+B)\bigg]R^3.
	\end{equation}
	
It is to be noted that the active gravitational mass also depends on the brane tension $\sigma$.

\subsection{Intermediate thin Shell}
The intermediate thin shell is a junction formed between the interior and exterior spacetimes. The shell is extremely thin but has a finite thickness. So one can consider $e^{-\lambda}\ll 1$. Under this thin shell approximation, the field equations \ref{eq6} - \ref{eq8} are modified as
\begin{equation} \label{eq16}
\frac{e^{-\lambda}\lambda^{\prime}}{r}+\frac{1}{r^2}=\bigg[\rho\bigg(1+\frac{6A}{\sigma}\bigg)+\frac{\rho^2}{2\sigma}+\frac{6B}{\sigma}\bigg],
\end{equation}

\begin{equation} \label{eq17}
-\frac{1}{r^2}=\bigg[\rho\bigg\{1+\bigg(\frac{1+2\omega}{\sigma}\bigg)2A\bigg\}+\frac{3\rho^2}{2\sigma}+\bigg(\frac{1+2\omega}{\sigma}\bigg)2B\bigg],
\end{equation}

\begin{equation} \label{eq18}
-\frac{\lambda^{\prime}\nu^{\prime}}{4}e^{-\lambda}-\frac{e^{-\lambda}\lambda^{\prime}}{2r}=
\bigg[\rho\bigg\{1+\bigg(\frac{1-\omega}{\sigma}\bigg)2A\bigg\}+\frac{3\rho^2}{2\sigma}+\bigg(\frac{1-\omega}{\sigma}\bigg)2B\bigg].
\end{equation}

The shell is composed of the Zeldovich stiff fluid with the EOS in the form $p=\rho$. Putting this EOS in the conservation equation on the brane, i.e. Eq. (\ref{eq9}), we obtain a relation between the metric potential $\nu$ and energy density of the shell $\rho$ as
\begin{equation}\label{eq19}
\rho=\rho_0 e^{-\nu}.
\end{equation}

Equation (\ref{eq17}) is a quadratic equation for $\rho$, considering the positive root of which we get the metric potential as
\begin{equation} \label{eq20}
e^{-\nu}=-\frac{J}{6G\rho_0}+\sqrt{\frac{J^2}{36G^2\rho_0^2}-\frac{H_2}{36\rho_0^2}-\frac{1}{3G\rho_0^2r^2}}.
\end{equation}

The field equations (\ref{eq16}) and (\ref{eq18}) can be solved to obtain
\begin{eqnarray}\label{eq21}
e^{-\lambda}=& &\frac{4\ln r}{3} + \bigg(\frac{J}{36}-\frac{F}{12}\bigg) \frac{r}{G}\sqrt{(J^2-12GH_2)r^2-12G} + \bigg(\frac{FJ}{12G}-\frac{H_1}{2}+\frac{H_2}{6}-\frac{J^2}{36G}\bigg)r^2\nonumber\\ & & +\frac{(F-\frac{J}{3})}{\sqrt{12GH_2-J^2}}\arctan\bigg[\frac{(\sqrt{12GH_2-J^2})r}{\sqrt{(J^2-12GH_2)r^2-12G}}\bigg],
\end{eqnarray}
where $F=1+\frac{6A}{\sigma}$, $G=\frac{1}{2\sigma}$, $H_1=\frac{6B}{\sigma}$, $H_2=2B\bigg(\frac{1+2\omega}{\sigma}\bigg)$, $J=1+2A\bigg(\frac{1+2\omega}{\sigma}\bigg)$, $K=1+2A\bigg(\frac{1-\omega}{\sigma}\bigg)$ and $H_3=2B\bigg(\frac{1-\omega}{\sigma}\bigg)$.

\subsection{Exterior region}
The exterior of the gravastar is assumed to obey the EOS, $p=\rho=0$, which means that the outside region of the shell is completely vacuum. In this case Eq. (6) reduces to
\begin{equation}\label{eq22}
e^{-\lambda}\bigg(\frac{\lambda^{\prime}}{r}-\frac{1}{r^2}\bigg)+\frac{1}{r^2}=\frac{6B}{\sigma}.
\end{equation}

The solution has the form
\begin{equation}\label{eq23}
e^{-\lambda}=1-\frac{2M}{r}-\frac{2B}{\sigma}r^2,
\end{equation}
where $M$ is integration constant.

This can be compared with a de Sitter solution in $(3+1)$ dimension and the corresponding line element can be written as
\begin{eqnarray} \label{eq24}
ds^2=\left(1-\frac{2M}{r}-\frac{\Lambda r^2}{3}\right)dt^2&-&\left(1-\frac{2M}{r}-\frac{\Lambda r^2}{3}\right)^{-1}dr^2-r^2(d\theta^2+\sin^2\theta d\phi^2).
\end{eqnarray}

Here the integration constant $M$ gives the total mass of the gravastar and $\Lambda$ is the effective brane cosmological constant given by $\Lambda=\frac{6B}{\sigma}$. Since we are considering a vacuum exterior, it can be argued from the RS model~\cite{Randall2} that the effective cosmological constant on the brane vanishes as a consequence of the fine tuning between the brane tension and the bulk cosmological constant. So, we claim $B=0=\Lambda$ for the vacuum exterior. Thus, as expected to be observed locally, the exterior solution reduces to the Schwarzschild solution
\begin{equation}\label{eq25}
ds^2=\left(1-\frac{2M}{r}\right)dt^2-\left(1-\frac{2M}{r}\right)^{-1}dr^2-r^2(d\theta^2+\sin^2\theta d\phi^2).
\end{equation}

\subsection{Matching Condition} 
In order to get the unknown constants we have adopted the matching condition of the metric functions at the junctions: (i) interior and shell ($r=R_1$), and (ii) shell and exterior ($r=R_2$). Now one can match $g_{tt}$ and $\frac{\delta g_{tt}}{\delta r}$ at $r=R_2$ to obtain the values of different constants, viz. $A = -504.9521017$, $B = 3.01787$ inside the shell. In order to study different features of gravastar we have taken the ratio of the matter densities of the shell and that of the core as $10^4\ (=\frac{\rho_0}{\rho_c})$, $\sigma=10^3 MeV/fm^3$ and $\omega=10^{-3}$. We have also considered the following numerical values: $m=3.75 M_\odot$, $R_1=10$ and $R_2=10.1$.

\section{Physical parameters of the model} \label{sec3}
In this section we shall be discussing some of the important physical parameters of the shell of gravastar.

\subsection{Pressure and matter density}
It has been considered that the shell is formed with ultrarelativistic matter of extremely high density. The EOS has been stated as $p=\rho$. Using \ref{eq17} we get the pressure as well as the matter density as follows
\begin{equation}\label{eq26}
\rho=p=\frac { \left( \sqrt {64 r^2 \left(\frac{k^8 \sigma^2}{16}+\frac{ \left( A\sigma-3\,B
 \right)  Y {k}^{4}}{4}+\frac{A^{2}}{4} Y^2 \right) \pi - {3k^8\sigma} \pi }+ \left(-4 Y A-2k^4\sigma \right) r \pi  \right)}{6 \pi k^4 r},
\end{equation}
where $Y= 2\omega+1$.

The variation of the energy density which is the same as the pressure of the shell along with the radial distance $r$ has been plotted in FIG. 1.

\begin{figure*}[thbp] 
\centering
\includegraphics[width=0.5\textwidth]{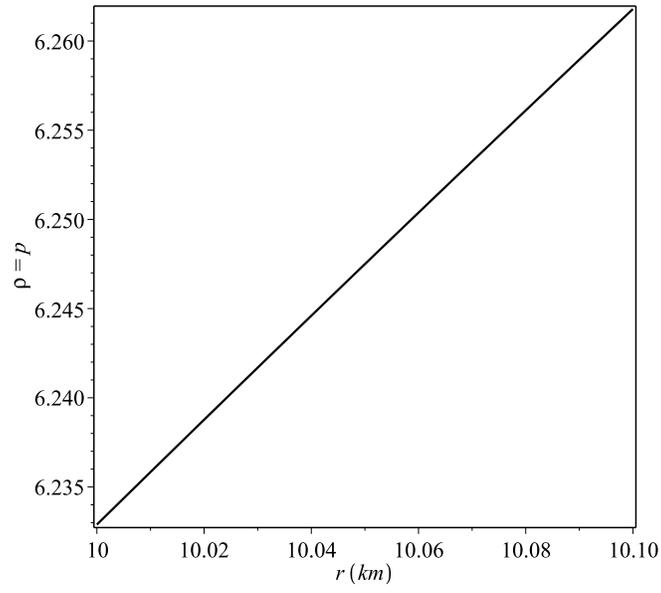}
\caption{Variation of the pressure or the matter density of the shell with respect to $r$.}
\end{figure*}

\subsection{Energy}
 To calculate the energy of the shell we integrate the following function which yields
 \begin{eqnarray}\label{eq27}
	E &=& 4\pi \int_{R+\epsilon}^{R}\rho^{eff}r^2 dr  \nonumber\\
	&=&\frac{2}{3 k^4} \left( {\frac { \left( 4\pi k^4 r^2 X -3 k^8\sigma+16 A^2 \pi r^2Y^2
 \right)^{3/2}}{12 \pi^\frac{1}{2} \left( k^4 X+
4A^2 Y^2 \right) }}-\frac{ r^3 \pi }{3}\left( 4 A Y+2 k^4\sigma \right)\right),
\end{eqnarray}
where $X=(k^4 \sigma^2+8 A\omega \sigma +4 A \sigma-24 B \omega-12 B)$ and $Y=(2\omega+1)$.

The variation of the energy of the shell along with the radial distance $r$ has been plotted in FIG. 2.

\begin{figure*}[thbp] 
\centering
\includegraphics[width=0.5\textwidth]{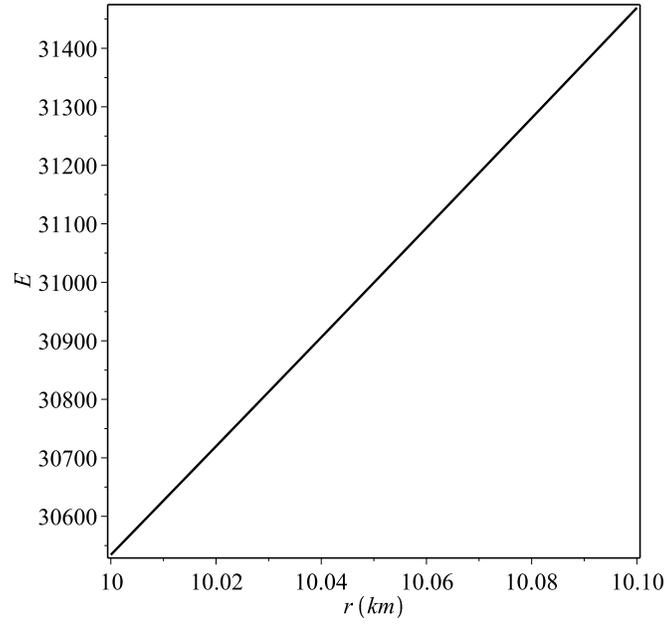}
\caption{Variation of the energy of the shell with respect to $r$.}
\end{figure*}

\subsection{Entropy}
The entropy is one of the most important parameters associated with a black hole. So, we
must compute the entropy for a gravastar too. The entropy of the gravastar on the brane
can be calculated using the following equation
\begin{equation}\label{eq28}
S=\int_R^{R+\epsilon} 4 \pi r^2 s(r) \sqrt{e^{\lambda}} dr.
\end{equation}

Here $s(r)$ is defined as the entropy density and can be written as
\begin{equation}\label{eq29}
 s(r)=\frac{\xi^2 k_B^2 T(r)}{4\pi\hbar^2}=\frac{\xi k_B}{\hbar}\sqrt{\frac{p}{2\pi}},
\end{equation}
where $\xi$ is a dimensionless constant. Considering the geometrical units, i.e. $G=1,\ c=1$, and also in the Planck units $k_B=1, \hbar=1$, Eq. (\ref{eq29}) yields the following relation
 \begin{equation}\label{eq30}
 s(r)=\xi\sqrt{\frac{p}{2\pi}}.
\end{equation}

Therefore, the entropy of the shell obtained as
\begin{eqnarray}\label{eq31}
S&=& 4\pi \xi \int_{R}^{R+\epsilon} r^2 \sqrt{\frac{p e^\lambda}{2\pi}}dr \nonumber\\
&=& 4\pi \epsilon r^2\bigg[\frac{4\ln r}{3} + \bigg(\frac{J}{36}-\frac{F}{12}\bigg) \frac{r}{G}\sqrt{(J^2-12GH_2)r^2-12G} + \bigg(\frac{FJ}{12G}-\frac{H_1}{2}+\frac{H_2}{6}-\frac{J^2}{36G}\bigg)r^2 \nonumber\\  &+&\frac{(F-\frac{J}{3})}{\sqrt{12GH_2-J^2}}\arctan\bigg\{\frac{(\sqrt{12GH_2-J^2})r}{\sqrt{(J^2-12GH_2)r^2-12G}}\bigg\}\bigg]^{-\frac{1}{2}}.
\end{eqnarray}
	
\begin{figure*}[thbp] 
	\centering
	\includegraphics[width=0.5\textwidth]{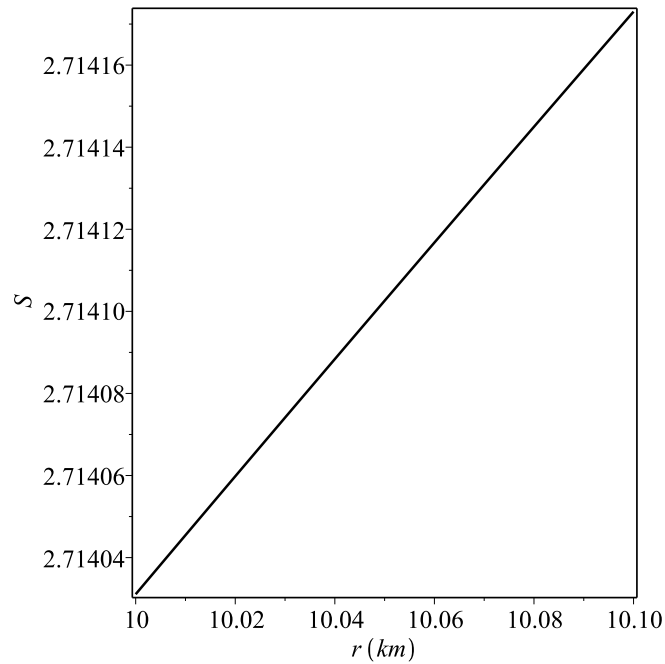}
	\caption{Variation of the entropy $(S)$ of the shell with respect to $r$.}
\end{figure*}

The variation of entropy of the shell along with the radial distance $r$ has been plotted in FIG. 3.

\subsection{Proper thickness}
The shell is considered to be extremely thin so that the phase boundaries are taken to be at $R$ and $R+\epsilon$, where $\epsilon \ll 1$, such that the phase boundary of the interior essentially is at $R$. Therefore, the proper thickness of the shell is computed to be
\begin{eqnarray}\label{eq32}
\ell=& &\int_{R}^{R+\epsilon}\sqrt{e^\lambda}dr =\epsilon\sqrt{e^\lambda}\nonumber\\
& &= \epsilon \bigg[\frac{4\ln r}{3} + \bigg(\frac{J}{36}-\frac{F}{12}\bigg) \frac{r}{G}\sqrt{(J^2-12GH_2)r^2-12G} + \bigg(\frac{FJ}{12G}-\frac{H_1}{2}+\frac{H_2}{6}-\frac{J^2}{36G}\bigg)r^2\nonumber\\ & & +\frac{(F-\frac{J}{3})}{\sqrt{12GH_2-J^2}}\arctan\bigg\{\frac{(\sqrt{12GH_2-J^2})r}{\sqrt{(J^2-12GH_2)r^2-12G}}\bigg\}\bigg]^{-\frac{1}{2}}.
\end{eqnarray}

\subsection{Surface Redshift}
We compute the surface redshift in order to check the stability of our gravastar model. This is given by
\begin{eqnarray}\label{eq35}
Z_{s}&=&-1+\frac{1}{\sqrt {g_{\it tt}}} \nonumber\\
&=&-1+\sqrt{\frac{1}{6\rho_{0}\pi k^4 r} \left(8\sqrt{\left(r^2 \left(\frac{k^8{\sigma}^2}{16}
 +\frac{1}{2}\left( A\sigma-3\,B \right)  X {k}^{4}+
 X^2 A^2 \right) \pi -\frac {3 k^8 \sigma}{64} \right) \pi}-2 \left(4X A+ k^4\sigma \right) r\pi  \right)},
\end{eqnarray}
where $X=\left( \omega+\frac{1}{2} \right)$.

\begin{figure*}[thbp] 
\centering
\includegraphics[width=0.5\textwidth]{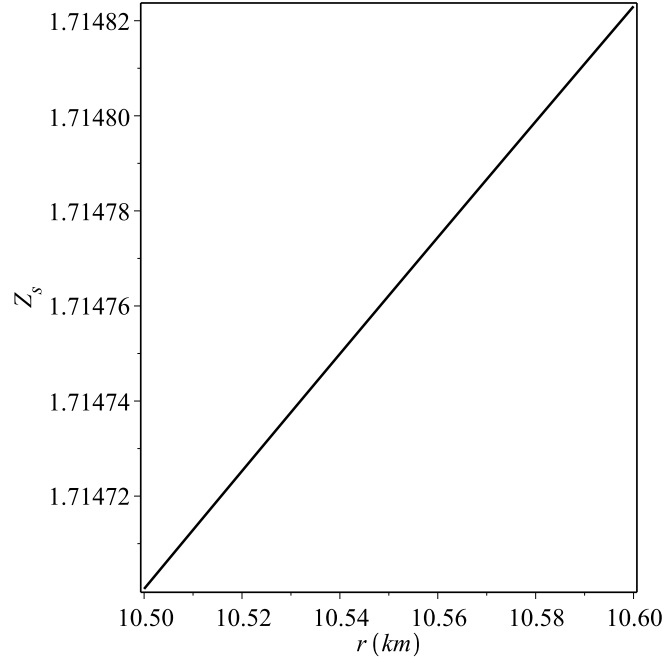}
\caption{Variation of the Surface redshift of the shell with respect to $r$.}
\end{figure*}

We have checked the variation of the surface redshift with respect to $r$ in FIG. 4. It has been observed that without cosmological constant the surface redshift $(Z_s)$ lies within the range $Z_s \leq 2$~\cite{Buchdahl1959,Straumann1984,Bohmer2006}. However, Bohmer and Harko~\cite{Bohmer2006} argued that $Z_s\leq5$ for the compact objects in the presence of cosmological constant. For our model, the surface redshift lies within $2$ at every point of the shell.

\section{Boundary condition} \label{sec4}
The gravastar comprises of the following three regions: (i) Interior (ii) Shell and (iii) Exterior. The shell connects the interior to the exterior region at the junction interface. The metric coefficients are continuous across the interface but the continuity of the first derivatives of the
metric coefficients is not confirmed. The Israel-Darmois~\cite{Darmois,Israel} junction conditions allow us to compute the intrinsic surface stress-energy at the junction in terms of the extrinsic curvature which connects the two sides of the thins shell geometrically. The intrinsic stress-energy, following the prescription of Lanczos~\cite{Lanczos}, turns out to have surface energy density and surface pressures as the temporal and spatial components respectively. These components are computed to be of the form
\begin{eqnarray}\label{eq33}
    \sigma & & =-\frac{1}{4\pi R}\bigg[\sqrt{e^\lambda}\bigg]_-^+ \nonumber\\& &
     =\frac{1}{4\pi R}\bigg[\sqrt{\left(1-\frac{2M}{R}\right)}- \sqrt{1-\bigg\{\frac{\rho_c}{3}\bigg(\frac{2\sigma+\rho_c}{2\sigma}\bigg)+\frac{2(A\rho_c+B)}{\sigma}\bigg\}R^2}\bigg],
	\end{eqnarray}

\begin{eqnarray}\label{eq34}
\mathcal{P} & =&\frac{1}{16\pi } \bigg[\bigg(\frac{2-\lambda^\prime R}{R}\bigg) \sqrt{e^{-\lambda}}\bigg]_-^+ \nonumber\\
&=&\frac{1}{8\pi R}\left[\frac{1-\frac{M}{R}}{ \sqrt{1-\frac{2M}{R}}}- \frac{1-2\bigg\{\frac{\rho_c}{3}\bigg(\frac{2\sigma+\rho_c}{2\sigma}\bigg)+\frac{2(A\rho_c+B)}
{\sigma}\bigg\}R^2}{\sqrt{1-\bigg\{\frac{\rho_c}{3}\bigg(\frac{2\sigma+\rho_c}{2\sigma}\bigg)+\frac{2(A\rho_c+B)}{\sigma}\bigg\}R^2}}\right].\label{eq70}
\end{eqnarray}

 The variation of the surface pressure of the shell has been shown in FIG. 5.

\begin{figure*}[thbp] 
\centering
\includegraphics[width=0.5\textwidth]{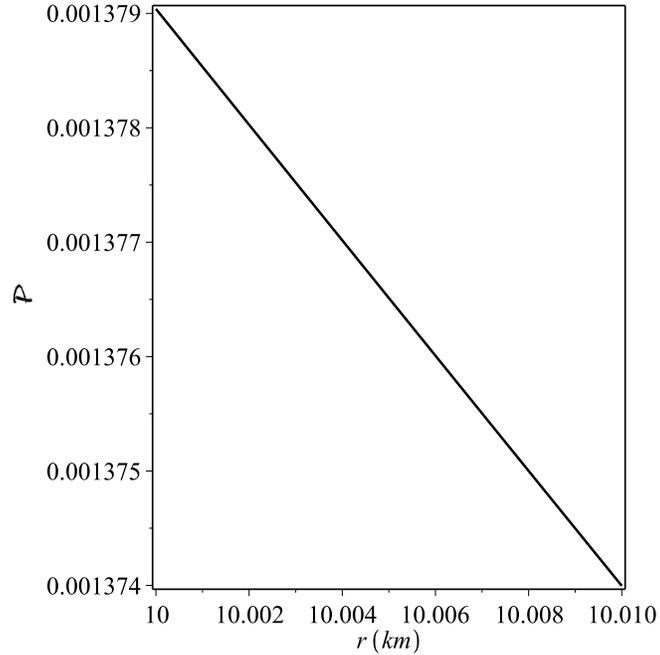}
\caption{Variation of the Surface pressure of the shell with respect to $r$.}
\end{figure*}

Now, the mass of the thin shell can be written as
\begin{equation}
m_s=4\pi R^2\sigma= R \bigg[\sqrt{\left(1-\frac{2M}{R}\right)}- \sqrt{1-\bigg\{\frac{\rho_c}{3}\bigg(\frac{2\sigma+\rho_c}{2\sigma}\bigg)+\frac{2(A\rho_c+B)}{\sigma}\bigg\}R^2}\bigg].\label{40}
\end{equation}

\section{Discussions and Conclusion} \label{sec5}
In this work we have studied gravastar under the framework of RS-2 brane gravity. The study of this model of gravitation
is found interesting not only in the context that it modifies the EFE but also the higher dimensions is
involved. Following the earlier work done on gravastar under the modified theory of gravity
models~\cite{Shamir,Das,Debnath2019a,Debnath2019b} a detail study have been done for the three different regions of gravastar under braneworld theory. 

Let us summarize some of the important physical properties of our study as follows:

{\textbf 1. Interior Region}:
Solving the TOV equation along with the EOS of the interior it has been found that the matter
density as well as the pressure remains constant in the interior and the solution is found
to be free from singularity. We have also calculated the active gravitational mass of the
interior and it is observed that it has an additional dependence on the brane tension.

{\textbf 2. Intermediate Thin Shell}: To solve the intermediate thin shell we have applied
the thin shell approximation and computed metric functions of it. The metric functions are found
to be modified due to the brane-world effects reflected in the dependence on the brane tension and the bulk EOS parameter.

{\textbf 3. Physical parameters of the Shell:} Various physical parameters associated with the shell have been computed and the behaviour is found to be modified due to the brane effects - both local and non-local. The details are provided below:

{ \textbf (i) Matter density:} We have calculated the matter density and the pressure of the shell and plotted it
against $r$ as shown in FIG. 1. The variation of matter density or pressure for the shell is found
to be positive and constantly increases as we move from the interior to the exterior surface.

{ \textbf (ii) Energy:} The energy of the shell has been obtained in Eq. (\ref{eq27}) and variation with respect to
radial parameter is shown in FIG. 2. The graph shows similar nature as the matter density of the shell,
which suggests regarding the physical acceptability of the model.

{\textbf (iii) Proper Length and Entropy:} We have also calculated the Entropy and the proper thickness
of the shell and the solution are found to be physically acceptable. The variation of entropy with the radial parameter has been plotted in FIG. 3 which shows a constantly increasing nature.

{\textbf (iv) Surface energy density and surface pressure:} Following the condition of Darmois and
Israel~\cite{Darmois,Israel} we have calculated the surface energy density and surface pressure and plotted surface pressure against the radial parameter (FIG. 4). The surface pressure remains positive throughout the shell and decreases as we move from the inner boundary of the shell to the outer which supports the formation and existence of the thin shell between the two spacetimes, i.e. interior and exterior.

{\textbf (v) Surface redshift:} We have checked the stability of the gravastar through
surface redshift analysis, for any stable model the value of surface redshift lies within 2~\cite{Buchdahl1959,Straumann1984,Bohmer2006}.
For our model we have found that our model is stable under surface redshift as can be noticed from FIG. 5.

Now the question arises regarding the possible existance and detection of the gravastar under
our present study. Though there is no direct evidences to detect garavastar but some of the
indirect ways have been discussed in literature~\cite{Sakai2014,Kubo2016,Cardoso1,Cardoso2,Abbott2016,Chirenti2016}.
The idea for possible detection of gravastar was first proposed by Sakai et al.~\cite{Sakai2014} through
the study of gravastar shadows. Another possible method for the detection of gravastar may be
employing gravitational lensing as suggested by Kubo and Sakai~\cite{Kubo2016} where they have claimed to have found gravastar microlensing effects of larger maximal luminosity compared to black holes of the same mass. According to Cardoso et al.~\cite{Cardoso1,Cardoso2}, the ringdown signal of $GW~150914$~\cite{Abbott2016} has been detected by interferometric LIGO detectors are most probably generated by objects without event horizon which might be garvastar, though it yet to be confirmed~\cite{Chirenti2016} (in this context a detailed review on gravastar can be found in the ref.~\cite{Ray2020}).

As a final comment, we can conclude that in the present paper a successful study has been done on gravastar under
the brane world theory of gravity. We have obtained a set of physically acceptable and non
singular solution of the gravastar, which immediately overcome the problem of the central singularity
and existence of event horizon of black hole. One can note that this work provides a general
solution of the gravastar under the framework of brane-world gravity without admitting
conformal motion unlike Banerjee et al.~\cite{Banerjee1}. The solution for the exterior metric is found to
be Schwarzschild type whereas they found the solution as Reissner-Nordstrom type. Analysing
all the results that we have obtained, we claim the possible existence of gravastar in brane world theory as obtained in Einstein's GR.

\section*{Acknowledgments}
SR, BM and SKT are thankful to the Inter-University Centre for Astronomy and Astrophysics (IUCAA),
Pune, India for providing the Visiting Associateship under which a part of this work was carried out.

\end{document}